
\input vanilla.sty

\magnification 1200

\hsize=16truecm\advance\hoffset by -.5cm
\vsize=23.5truecm

\parskip=0pt plus 1pt
\advance\baselineskip by 6pt
\overfullrule=0pt

\font\gross=cmcsc10
\font\Titel=cmr10 scaled \magstep 2
 1
\font\eightrm=cmr8
\font\eightsl=cmsl8

\def\norm#1{\Vert #1\Vert}
\def\R{\hbox{{\rm I}\kern -0.2em {\rm R}}}
\def\mc{measure cone}
\def\tr{$\bigl(V,V^+,e\bigr)$}
\def\md{minimal decomposition}

\topinsert

\line{\hss {\eightsl To
appear in: Letters in Mathematical Physics (1998).}
}

\vskip 1.5truecm 
\endinsert

\centerline{\Titel Orthogonality and Disjointness} 
\bigskip
\centerline{\Titel in Spaces of Measures}

\vskip 1truecm

\centerline{\gross Paul Busch}
\centerline{Department of Mathematics}
\centerline{The University of Hull}
\centerline{Hull HU6 7RX, United Kingdom}

\bigskip
\centerline{\eightrm November 1997}
\vskip 1truecm

{\advance\baselineskip by -6pt
{\narrower\noindent\eightsl
The disjointness of measures and their Hahn-Jordan decomposition
are instances of the general notion of minimal decomposition in
base normed spaces. The mixing distance, a specification of
a novel concept of angle in real normed vector spaces, is applied to provide
a geometric interpretation of disjointness as orthogonality.
\par}

\vskip 12pt
{\narrower\noindent{\eightrm Key words}: {\eightsl  Ordered vector
space, base norm, minimal decomposition, measure cone, direction 
distance, mixing distance.}\par}
}

\vskip 8pt
\noindent{\bf 0. Introduction}
\vskip 3pt
\noindent
 The convex and metric structures underlying probabilistic physical
theories are generally described in terms of base normed 
vector spaces. According to a recent proposal, the purely geometrical
features of these spaces are appropriately represented in terms of the
notion of {\sl measure cone} and the {\sl mixing distance} [1], a
specification of the novel concept of {\sl direction distance} [2]. It
turns out that the base norm is one member of a whole characteristic
family of {\sl mc-norms} from which it can be singled out by virtue of
a certain orthogonality relation.  The latter is seen to be closely
related to the concept of {\sl minimal decomposition}. These
connections suggest a simple geometric interpretation of the familiar
notion of the disjointness of (probability) measures and the 
Hahn-Jordan decomposition of measures 
which has been addressed briefly in [1] and will be
elaborated here. The results
obtained give an indication of the extent to which a general measure
cone admits measure theoretic interpretations. For the sake of a
self-contained presentation, the basic structures are briefly
reviewed first. 

\vskip 8pt
\noindent
{\bf 1. Direction Distance and Mixing Distance in Measure Cones}
\vskip 5pt

\subheading{1.1}
In a fundamental investigation [2] Ernst Ruch proposed an
interesting extension of Felix Klein's geometry programme for affine geometries
based on normed real vector spaces $V$. Defining the figure of an 
{\sl (oriented) angle}
as an ordered pair $\bigl([x],[y]\bigr)$ of directions (rays, $[x]:=
\{\lambda x|\lambda > 0\}$ for $x\neq 0$), two angles are called
{\sl norm-equivalent}, $\bigl([x],[y]\bigr)\ \sim\ 
\bigl([x'],[y']\bigr)$, if there exists a linear map that sends
$[x]\cup [y]$ onto $[x']\cup [y']$ 
such that corresponding pairs of points are equidistant:
$\parallel \alpha x_0 - \beta y_0\parallel\ =\
\parallel \alpha x^{\prime}_0 - \beta y^{\prime}_0\parallel$ 
for all $\alpha,\beta\in\R^+$.
(Here $x_0:=x/{\parallel x\parallel}$, etc.). 

The {\sl direction distance} $d[x/y]$ (from $x$ to $y$) is defined as 
the family of distances $||\alpha x_0-\beta y_0||$. This is succinctly
summarized in terms of the map
$$
d:\bigl([x],[y]\bigr)\mapsto d[x/y],\ \ 
d[x/y]:\R^+\times\R^+ \to \R^+,\quad (\alpha,\beta)\mapsto 
\norm{\alpha x_0 -\beta y_0},\eqno(1)
$$
The direction distance 
induces an ordering of the classes of norm-equivalent angles via
$$
d[x/y] \succ d[x^\prime /y^\prime ]\ :\iff\ \forall \alpha ,\beta\in\R^+:
\norm{\alpha x_0 -\beta y_0} \geq\norm{\alpha x^\prime_0 -\beta y^\prime_0}.
\eqno(2)
$$

\subheading{1.2.\ Definition}
Let $V$ be a normed real vector space. Elements
$z,z'\in V$ are called ({\sl norm-}){\sl orthogonal}, 
$z\perp z'$, if the following
condition is satisfied:
$$
\norm{\alpha z_0 -\beta z_0^{\prime}}\ \ =\ \ 
\norm{\alpha z_0 +\beta z_0^{\prime}}\qquad 
\forall\alpha,\beta\in\R.\eqno(3)
$$
This notion of orthogonality is based on the equivalence of an angle
$\left([z],[z']\right)$ and its (planar) complement
$\left([z],[-z']\right)$: $d[z/z']=d[z/-z']$.

\subheading{1.3}
With the above definitions the concept of angle is a straightforward
generalization of the canonical notion of angle in inner product
(pre-Hilbert) spaces. 
The inner product metric of angles is characterized by 
the requirement that the direction distance is a symmetric function of its 
arguments, or equivalently, that the ordering of angles (2) is a
total ordering, or equivalently, that the group of linear isometries
acts transitively on the norm unit sphere [2]. 
In an inner product space the orthogonality condition (3) is equivalent
to $||z_0-z_0'||=||z_0+z_0'||$ and to the vanishing of the inner
product between $z_0$ and $z_0'$.
In the present context another
family of normed vector spaces will be considered
which are characterized by the fact that only a reduced symmetry 
is present: a transitive action of a group of isometries is given only 
on a certain subset of the norm unit sphere.

\subheading{1.4. Definition} 
                A triple $(V,V^+,e)$ is a measure cone if the following
                postulates are satisfied:

\item{(a)\  }   $V$ is a real vector space with convex, generating cone
                $V^+$ ($V\ =\ V^+-V^+$).

\item{(b)\  }   $e:V\rightarrow\R$ is a linear functional, 
                called {\sl charge}, that is
                strictly positive,
                $$z\in V^+\ \Longrightarrow\ \big\{e(z)\geq 0\ {\text{and}}
                \ e(z)=0\Leftrightarrow z=0\big\}.\eqno(4)
                $$
                It follows that the charge $e$ admits a
                decomposition $e=e_+-e_-$ of $e$  
                into a difference of nonlinear, positive functionals 
                $e_{\pm}$, where
$$
\eqalign{
e_+:\ V\ \to\ \R^+,\ \ 
z\ &\mapsto\  e_+(z)\ 
:=\ \inf\bigl\{e(x)\vert\ x\in V^+,\ x-z\in V^+\bigr\},\cr 
e_-:\ V\ \to\ \R^+,\ \ 
z\ &\mapsto\ e_-(z)\  
:=\ \inf\bigl\{e(y)\vert\ y\in V^+,\ z+y\in V^+\bigr\}.\cr}
\eqno(5)
$$
\item{ }        Further, it is required that $e$ marks the cone contour:
                $$z\in V^+\ \ \Longleftrightarrow\ \ e(z)\ =\ e_+(z).
                \eqno(6)$$ 

\flushpar A measure cone \tr\ is said to be a \mc\ with {\sl \md} 
if in addition the following postulate is satisfied: 

\item{(c)\  }   To any $z\in V$ there exists a decomposition
                $z=z_+-z_-$, $z_+,\ z_-\in V^+$ such that 
                the following  holds: $e(z_+)=e_+(z),\ e(z_-)=e_-(z)$.
                Any decomposition of $z$ with this property is called
                a minimal decomposition of $z$.
\flushpar 
A real vector space $V$ equipped with a \mc\ \tr\ (with \md) will be called
mc-space (with \md).

All known physically relevant examples of measure cones are
equipped  with a minimal decomposition.
Hence in the sequel the term measure cone will be taken to always 
include the existence of a minimal decomposition.

\subheading{1.5}
The structure of a measure cone and its  properties will be briefly 
summarized; details and proofs can be found in [1]. 
The set $V^+$ is a proper (convex) cone so that
$V$ becomes an ordered vector space via 
$z\geq z'\quad :\Leftrightarrow\quad z-z'\in V^+$.
The strict positivity of the charge functional $e$ ensures that the 
intersection $K$ of the hyperplane $\{z\in V|e(z)=1\}$ with $V^+$
is a base of the convex cone $V^+$. 
In a measure cone with minimal decomposition
the cone contour condition (6) is a consequence of the
strict positivity of $e$. 

Any vector space $V$ associated with a measure
cone can be equipped with a norm. In general a triple \tr\ consisting
of a real vector space $V$, a convex generating cone $V^+\subset V$ and
a linear functional $e$ is a measure cone if and only if there exists a norm
$\parallel\cdot\parallel$ marking the cone contour in the following sense:
$$
z\in V^+\ \ \Longleftrightarrow\ \ e(z)=\norm z.   \eqno(7)
$$
In particular, the following is a norm of this type, called 1-norm:
$$
{\norm z}_1\ \ :=\ \ e_+(z)+e_-(z),  \eqno (8)
$$
This norm turns out to coincide with the
base norm, defined as the Minkowski functional of the (absorbing, balanced,
radially bounded) set $B:=\text{co}(K\cup -K)$, the convex
hull of $K\cup -K$; thus, 
${\norm z}_1\  =\ \text{inf}\{\lambda\geq 0|z\in\lambda B\}$ (cf.\ [3]).

Measure theoretic considerations suggest (cf. [1]) to define a family 
of norms of the form 
$$
\norm z \ := \ \Cal P\bigl(e_+(z),e_-(z)\bigr)\eqno (9)
$$
which in addition are required to mark the cone contour in the sense
of Eq.\ (7). Such norms, called {\sl mc-norms}, have been
completely characterized [1].
A mapping of the form (9) is a norm if and only if
the function $\Cal P$, defined on $\R^+\times\R^+$, has an extension
to a norm on $\R^2$ and possesses the symmetry $\Cal P(a,b)=\Cal P(b,a)$.
Any such norm is an mc-norm if and only if $\Cal P$ is strictly monotonic
in the following sense:
$$
\delta > 0\ \Longrightarrow\ \Cal P(a+\delta,b+\delta) > \Cal P(a,b)
\qquad \text{for all}\ (a,b)\in\R^+\times\R^+.
$$
An mc-space equipped with an mc-norm is called an mc-normed space.
The 1-norm is characterized as the mc-norm with  the smallest norm unit ball:
$$
|e(z)|\ \ \leq\ \ \norm{z}\ =\ \Cal P\bigl(e_+(z),e_-(z)\bigr)\ \ 
\leq\ \ \norm{z}_1.
$$

\subheading{1.6}
The {\sl mixing distance} is defined as the restriction of the direction
distance  to pairs of directions in the
positive cone $V^+$ of an mc-normed space $V$. 
All mc-norms associated with a given measure cone are
equivalent in that they characterize the same ordering of
angles between pairs of 
directions in $V^+$ (Theorem 3.1 of [1]). 
Nevertheless the 1-norm is singled
out by the notion of orthogonality [Proposition 2.2 below].

\subheading{1.7}
The use of a measure theoretic terminology in the context of mc-spaces 
is motivated by the following considerations.
A base normed space $\bigl(V,\norm{\cdot}_1\bigr)$ and its dual 
order unit space $(V',e)$  provide a general probabilistic framework 
that was found useful for physical applications and is
referred to as a {\sl statistical duality} (cf. [4] and 
references therein). The set $K\subset V$ is taken to be the
convex set of statistical states of a physical system. 
The positive part $E:=[o,e]$ of the order unit interval
of $V'$ is a partially ordered, convex set 
of positive linear functionals on $V$. $E$ is
called the set of {\sl effects} as it represents the totality of yes-no
propositions that can be made about a given physical system. 
$E$ is equipped with a complement operation,
$a\mapsto a':=e-a$. Now the elements $z$ of $V^+$ (or of $K$), 
considered as linear functionals
on $V'$ via $z(a):=a(z)$, act as positive, additive [$z(a+b)=z(a)+z(b)$
whenever $a+b\in E$] (and normalized, $z(e)=1$, if $z\in K$) 
functions on $E$, representing thus (probability) measures in a 
generalized sense. 

\subheading{1.8. Definition} Elements $a,b\in E$ are {\sl
(weakly) orthogonal}, $a\perp b$, if their sum is in $E$
again, that is, if $b\leq a'$.
 
The map $a\mapsto a'$ is not an orthocomplementation:
one has $(a')'=a$ and $a\le b\Rightarrow b'\le a'$ but not in general
$a\land a'=o$. [For example, if $a=\frac 12 e$ then $a'=a$]. Here
$a\land b$ denotes the greatest lower bound of $a,b$. Effects $a,b$
are called {\sl disjoint} if $a\land b=o$. Hence weak orthogonality of
effects does not in general imply their disjointness.

\vskip 8pt
\noindent
{\bf 2. Orthogonality in a Measure Cone}
\vskip 5pt

\subheading{2.1}
The ordered set  of classes of equivalent angles induced by the direction 
distance has a smallest and a largest element represented by the pairs
$\left([x],[x]\right)$ and $\left([x],[-x])\right)$, respectively.
Orthogonality for a pair of directions in the cone $V^+$
is equivalent to maximality of the corresponding angle: 
for $x,y\in K$, $x\perp y$ iff
$d[x/y](\alpha,\beta)=\alpha +\beta = d[x/-x]$
(for all $\alpha,\beta$). The following Proposition gives a geometric
interpretation of the minimal decomposition in relation to 
the 1-norm [1].

\subheading{2.2. Proposition} Let \tr\ be a \mc. 
\item{(1)}  A decomposition $z = z_+ -z_-$ of $z \in V$
            with $z_+,z_-\in V^+$ is a minimal 
            decomposition iff it is an orthogonal decomposition 
            with respect to the 1-norm [that is, $z_+\perp_1 z_-$] iff
	    $\norm{z_+-z_-}_1=\norm{z_++z_-}_1$.

\item{(2)}   Let $\norm\cdot_{\Cal P}$ be a norm of the form (9) such that 
             $x\perp y$
             for at least one pair $x,y \in K$. Then $\norm\cdot_{\Cal P} =
             \norm\cdot_1$. Thus the 1-norm is the only mc-norm 
             allowing for a nonempty orthogonality relation on
             $V^+\times V^+$.

\subheading{2.3. Definition} Let $V$ be an mc-space equipped with the
1-norm, $E=[o,e]\subset V'$ the associated set of effects. Elements
$x,y\in V^+$ are called {\sl disjoint}, $x\dot\perp y$, if there
exists $a_x\in E$ and $a_y\in E$ such that
$a_x\perp a_y$ and $a_x(x)=e(x)$,  
$a_y(y)=e(y)$ (and hence $a_x(y)=0=a_y(x)$).

\noindent 
Note that $a_y$ can be chosen to be $a_y=a^{\prime}_x$. In fact
$x\dot\perp y$ iff there exists $a_x\in E$ such that $a_x(x)=e(x)$,
$a_x(y)=0$.
This definition captures the idea that disjoint pairs of (positive) 
measures are carried by weakly orthogonal effects.

\subheading{2.4. Theorem} Let $V$ be an mc-space equipped with the 1-norm.
For any pair $x,y\in V^+$, $x\perp y$ if and only if $x\dot\perp y$.

\noindent
{\sl Proof.} The following facts will be employed. First, the norm
unit ball of the order unit space $V'$ coincides with the order unit
interval $[-e,e]$ [3]. Next, as a consequene of the Hahn-Banach theorem,
for $z\in V$ there exists $a_z\in [-e,e]$ such that $a_z(z)=\norm{z}_1$.
\newline\noindent
Let $x\perp y$ for some $x,y\in V^+$. By Proposition 2.2, $z=x-y$ is a
minimal decomposition of $z$. Hence, 
$a_z(z)=a_z(x)-a_z(y)=\norm{z}_1=e(x)+e(y)$
for some $a_z,\ -e\leq a_z\leq e$. It follows that $a_z(x)=e(x)$ and 
$a_z(y)=-e(y)$. Define $a_x\in E=[o,e]$ as $a_x:=\bigl(e+a_z\bigr)/2$,
then $a_x(x)=e(x)$, $a_x(y)=0$ and therefore $x\dot\perp y$.
\newline\noindent
Conversely, let $x\dot\perp y$ for some $x,y\in V^+$. 
Taking $a_x$ as in Definition 2.3,  
define $a:=2a_x-e\in [-e,e]$. It follows that
$a(x-y)\ =\ 2a_x(x-y)-e(x-y)\ =\ e(x)+e(y)\ =\ \norm{x+y}_1.$
On the other hand, one generally has
$a(x-y)\ \leq\ \norm{x-y}_1\ \leq\ \norm{x+y}_1.$
Therefore,
$\norm{x-y}_1\ =\ \norm{x+y}_1\ =\ e(x)+e(y)$,
which is to say that $z=x-y$ is a minimal decomposition and hence $x\perp y$.
This completes the proof.

This theorem generalizes the measure theoretic fact that the disjointness of
positive measures can be characterized as an orthogonality
relation with respect to the total variation norm.
Similarly it comprises the quantum mechanical case where two density
operators are disjoint if the projections onto their ranges are mutually
orthogonal, and this property is equivalent to the orthogonality of
these density operators with respect to the trace norm.

\subheading{2.5}
In the context of measure theory and in the case of quantum
mechanical density operators the minimal  decomposition
into a difference of orthogonal positive components is known to be
unique (the Hahn-Jordan decomposition or the spectral decomposition
according to the positive and negative parts of the spectrum).
However there exist measure cones which do admit 
non-unique minimal decompositions. 
For example [1], consider the measure cone generated by $V=\R^3$ and 
$K$ the square in the plane $z=1$ with vertices $(\pm 1,\pm 1,1)$.
Then the vector $(1,0,0)$ has the following family of minimal
decompositions:
$
(1,0,0)\ =\ {1\over{2}}(1,\alpha,1)\ -\ {1\over{2}}(-1,\alpha,1)$, 
 $ -1\leq\alpha\leq 1.
$
 This is verified by observing that the vectors $(1,\alpha,1)$, 
$(-1,\alpha,1)\ \in \ K$ are orthogonal with respect to the 1-norm and then
applying Proposition 2.2. 
A geometric uniqueness criterion for minimal decompositions
in terms of an intersection property
has been given in [5].   
Uniqueness also holds in an mc-space $V$ which is a vector
lattice under the ordering induced by its positive cone $V^+$
(classical case), since in  that case the 
unique lattice theoretical 
orthogonal decomposition coincides with the minimal decomposition [6].
Here a measure theoretic condition will be
derived which covers classical and quantum probabilistic
theories as well as intermediate cases.
In general this uniqueness can be ascertained 
for any mc-space with the property that
to every vector there corresponds a kind of support functional in
the set of effects.

{
\subheading{2.6. Definition} An mc-space $V$ (equipped with the 1-norm)
is said to have a {\sl support family} if its dual space $V'$
contains a family of {\sl support functionals} $s_z$, $z\in V$,
characterized by the following conditions:
$$
\eqalignno{
s_z&\in [o,e],&(S1)\cr
\norm{z}_1&=s_z(z_++z_-),&(S2)\cr
a\in [o,e],\ a(z_++z_-)=\norm{z}_1\ &\Rightarrow\ s_z\leq a,&(S3)\cr
x_1,x_2\in V^+\ &\Rightarrow \ s_{x_1-x_2}\leq s_{x_1+x_2},&(S4)\cr
x_1,x_2\in V^+,\  
s_{x_1} \perp s_{x_2}\ &\Rightarrow\ s_{x_1}\land s_{x_2}=o.&(S5)\cr
}
$$
Here $z=z_+-z_-$ denotes a minimal decomposition of $z$.
}

\subheading{2.7} 
It will be shown below that in an mc-space $V$ with support family there
exists only one minimal decomposition for each $z\in V$. At the present
stage it is appropriate to observe that 
the defining conditions for the support functionals are
independent of the particular choice of a minimal decomposition (if
there were a choice): Let
$z=z_+-z_-=z^{\prime}_+-z^{\prime}_-$ denote two minimal
decompositions of $z$.  Then $s_z(z_+)+s_z(z_-)=e(z_+)+e(z_-)$, so that in
view of $s_z(z_{\pm}) \leq e(z_{\pm})$ it follows that $s_z(z_{\pm})=
e(z_{\pm})$. Similarly, $s_z(z^{\prime}_{\pm})=e(z^{\prime}_{\pm})$.
But $e(z_{\pm})= e(z^{\prime}_{\pm})$, hence
$s_z(z_{\pm})=s_z(z^{\prime}_{\pm})$. 

Property $(S1)$ is motivated by the fact that the support of a measure
is represented by a characteristic function and hence an extreme effect;
similarly, the support of a self-adjoint (trace class) operator is
given by a projection operator, which is an extreme effect. $(S2)$,
$(S3)$ and $(S4)$ capture the idea that a support is a minimal carrier.
In particular it follows from $(S2)$, $(S3)$ that for any $z\in V$ the
support $s_z\in E$ is unique; so there exists only one map
$z\mapsto s_z$.
$(S5)$ is a  strengthening of the weak orthogonality of two
support effects into disjointness.

\subheading{2.8. Proposition} Let
$V$ be an mc-space with support family.  Then for $z\in V$, $x_1, x_2\in
V^+$ the following statements hold: 
$$ \leqalignno{ 
z=0\ \ &\Longleftrightarrow \ \ s_z=o;&(\text{i})\cr
x_1\le x_2\ \ &\Longrightarrow\ \ s_{x_1}\le s_{x_2};
&(\text{ii})\cr
s_{x_1}\leq s_{x_2}\ \ &\Longleftrightarrow\ \ s_{x_2}=s_{x_1+x_2};
&(\text{iii})\cr
x_1\dot\perp x_2\ \ &\Longleftrightarrow\ \ 
s_{x_1}\perp s_{x_2}\ \Longrightarrow\ \ s_{x_1+x_2}\le s_{x_1}+s_{x_2}.
&(\text{iv})\cr} 
$$

\noindent
{\sl Proof.}\ \ 
Ad (i): If $s_z=o$ then by $(S2)$: $z=0$. If $z=0$ then $z_+=z_-=0$ and 
so for $a=o$: $a(z_++z_-)=||z||_1=0$, hence by $(S3)$ $s_z\le a$ and so
by $(S1)$, $s_z=o$. 
\newline
Ad (ii): Assume $x_1\le x_2$. Then for $a\in [o,e]$, $a(x_2)=0$ implies
$a(x_1)=0$. For $a=e-s_{x_2}$ one has $a(x_2)=0$, hence
$s_{x_2}(x)=e(x)$. By $(S3)$, $s_{x_1}\le s_{x_2}$.\newline
Ad (iii): The relations
$s_{x_1}\le s{x_1+x_2}$, $s_{x_2}\leq s_{x_1+x_2}$ follow from (ii). 
Let $s_{x_1}\leq
s_{x_2}$, then $s_{x_2}(x_1+x_2)=\norm{x_1+x_2}_1$, so by ($S3$),
$s_{x_1+x_2}\leq s_{x_2}$ and therefore $s_{x_2}=s_{x_1+x_2}$. The
converse implication is obvious in view of $s_{x_1}\le s_{x_1+x_2}$.\newline
Ad (iv): 
Let $x_1\dot\perp x_2$. Thus there is $a_{x_1}\in[o,e]$ such that
$a_{x_1}(x_1)=e(x_1)$ and $a_{x_1}(x_2)=0$. Then for $a_{x_2}=e-a_{x_1}$
one has $a_{x_2}(x_2)=e(x_2)$, $a_{x_2}(x_1)=0$. So by $(S3)$, $s_{x_1}
\le a_{x_1}$, $s_{x_2}\le a_{x_2}$. Therefore $s_{x_1}+s_{x_2}\le
a_{x_1}+a_{x_2}=e$, that is, $s_{x_1}\perp s_{x_2}$.\newline
Let $s_{x_1}\perp s_{x_2}$. Then $s_{x_1}(x_1)=e(x_1)$, so
$s_{x_2}(x_1)=0$, and also $s_{x_2}(x_2)=e(x_2)$, so $s_{x_1}(x_2)=0$.
This is to say that $x_1\dot\perp x_2$. It also follows that 
$s_{x_1}+s_{x_2}(x_1+x_2)=e(x_1+x_2)$, so by $(S3)$, $s_{x_1+x_2}
\le s_{x_1}+s_{x_2}$.
This completes the proof.

\subheading{2.9. Remark} It is straightforward to see that
$s_x=s_{\lambda x}$ for $x\in K$, $\lambda\ne 0$, and $s_z=s_{z_++z_-}$
for $z\in V$. Hence the set of supports $s_x$ of elements $x\in K$
exhaust the whole family of supports. For any $x\in K$, define
$F_x:=\{y\in K|s_y\le s_x\}$. These sets are faces of the convex set
$K$, and the map $s_x\mapsto F_x$ is an order isomorphism where the
set of faces $F_x$ is ordered by set inclusion. Furthermore, for any two
supports $s_x,s_y$ the supremum exists and is $s_x\lor s_y=s_{x+y}
=s_{\frac 12 x+\frac 12 y}$. So $F_x\lor F_y=F_{\frac 12 x+\frac 12 y}$.
In the classical and quantum cases, the support family is given by the
extreme effects, that is, a Boolean algebra of sets underlying the
classical theory, and the orthocomplemented lattice of projections (closed
subspaces) of the underlying Hilbert space, respectively;
and these lattices are isomorphic
to the corresponding lattices of (closed) faces of the respective sets $K$
of probability measures or density operators.

\subheading{2.10. Theorem} Let $V$ be an mc-space (equipped with the 1-norm)
with support family. Then the minimal decomposition 
$z=z_+-z_-$ of any element $z\in V$ is uniquely determined.

\noindent
{\sl Proof.} Let $z=z_+-z_-=z^{\prime}_+-z^{\prime}_-$. It follows that
$z_+-z^{\prime}_+=z_--z^{\prime}_-$. By $s_{z_{\pm}}=s_{z'_{\pm}}$ and
Proposition 2.8. (iii) one has $s_{z_{\pm}+z'_{\pm}}=s_{z_{\pm}}=s_{z'_{\pm}}$.
Further, by ($S4$), $s_{z_{\pm}-z'_{\pm}}\leq s_{z_{\pm}+z' _{\pm}}$.
Therefore, $s_{z_{\pm}-z'_{\pm}}\leq s_{z_+}$ and $\leq s_{z_-}$. Now
$s_{z_+}\perp s_{z_-}$ [by Proposition 2.8. (iv)], so
that finally ($S5$) yields $s_{z_+-z'_+}=s_{z_--z'_-}=o$,
that is, by Proposition 2.8. (i),  
$z_+=z'_+$, $z_-=z'_-$. This completes the proof.

\subheading{3. Orthogonality-preserving mc-endomorphisms}

\subheading{3.1} The mappings on an mc-space $V$ which respect the structure
of the \mc, the mc-endomorphisms, are the linear, charge-preserving,
positive mappings on $V$. These mappings are contractions
with respect to any mc-norm. In particular, they lead to decreasing mixing
distance. Conversely, any linear, charge preserving
contraction is necessarily a positive mapping [1]. With the above results
it is straightforward to verify the following statement.

\subheading{3.2. Proposition} Let $V$ be an mc-space equipped with the
1-norm. The orthog\-onal\-ity-preserving mc-endomorphisms
on $V$ are exactly the positive isometries on $V$.

\noindent
{\sl Proof.} Let $\Phi$ be an orthogonality-preserving mc-endomorphism.
Then
$$\align
\norm{\Phi z}_1 &=e(\Phi z_+)+e(\Phi z_-)
\qquad [\text{$\Phi$ positive, orthogonality-preserving}]\cr
&=e(z_+)+e(z_-)\qquad [\text{$\Phi$ charge-preserving}]\cr
&=\norm{z}_1\cr
\endalign
$$
Conversely, assume $\Phi$ to be a positive isometry, then for a
minimal decomposition of $z$, $z=z_+-z_-$ one has
$$
e(\Phi z_+)+e(\Phi z_-)=e(z_+)+e(z_-)=||z_+-z_-||_1=||\Phi z_+-\Phi z_-||,
$$
so that $\Phi z=\Phi (z_+)-\Phi(z_-)$
is a minimal decomposition as well. By Proposition 2.2 this is to say
that any orthogonal pair $x,y\in V^+$ is mapped onto an orthogonal pair.
This completes the proof.

\subheading{3.3}
The last result can be elaborated into an operational
characterization of the (ir)reversi\-bility of a statistical state
transformation 
if the set $K$ is interpreted as a set of statistical states of a physical 
system and the mc-endomorphisms are interpreted as stochastic operators
on $V$ [7]. A particularly interesting problem is to specify those
mc-spaces for which the following implication is always true: for
states $x,y,x',y,\in K$, $d[x/y] \succ d[x'/y']$ implies that there
exists a linear state transformation $\Phi$ such that 
$(x',y')=(\Phi x,\Phi y)$. Partial answers are reviewed in
[1] and [2], and the general case in the context of
classical probability theory is demonstrated in [7].
It is also known that the above implication does not hold in general.
Still  a particular special case pertains to arbitrary mc-spaces.

\subheading{3.3. Proposition}
Let $V$ be an mc-space equipped with the 1-norm. Let $x,y\in K$,
$x\perp y$. Then  $d[x/y] \succ d[x'/y']$.
Take $a_x,a_y$ (as in Definition 2.3) so that 
$a_x+a_y=e$, $a_x(x)=a_y(y)=e(x)=e(y)=1$. Then $\Phi : z\mapsto
a_x(z)x'+a_y(z)y'$ is an mc-endomorphism, and 
$\Phi x=x'$, $\Phi y=y'$.

\subheading{4. Conclusion}

\noindent
The structure of a measure cone (equivalently, a base normed space)
with minimal decomposition allows for a natural geometric
interpretation of the disjointness of measures defined in terms of the
minimal decomposition. Components of a minimal decomposition are
mutually orthogonal (Proposition 2.2). Furthermore, orthogonal measures
are carried by weakly orthogonal effects (Theorem 2.4). In a \mc\ with
support family the minimal decompositions are unique (Theorem 2.10).
Finally the notion of orthogonality yields a characterization of {\it
isometric} mc-endomorphisms, i.e., reversible state transformations in
the sense of [8]. In the case of quantum mechanical systems it has
proved very useful in deriving the general form of isometric state
transformations [9].  In this way the concept of orthogonality
developed here is found to provide some insight into the
interpretation of (ir)reversibility in the sense of decreasing mixing
distance. 


\vskip 8pt
\noindent
{\bf References}
\vskip 5pt

\item{\ 1.} {\gross P.\ Busch, E.\ Ruch}: {\sl The Measure Cone -- 
            Irreversibility as a Geometrical Phenomenon}. Int.\ J.\ Quant.\ 
            Chem. {\bf 41} (1992) 163-185. 

\item{\ 2.} {\gross E.\ Ruch}: {\sl Der Richtungsabstand}.
            Acta Applic. Math. {\bf 30} (1992) 67-93.

\item{\ 3.} {\gross E.\ M.\ Alfsen}: {\sl Compact Convex Sets and Boundary
            Integrals}. Springer, Berlin, 1991.

\item{\ 4.} {\gross W.\ Stulpe}: {\sl Conditional Expectations, Conditional
            Distributions, and A Posteriori Ensembles in Generalized 
            Probability Theory.} Int. J. Theor. Phys. {\bf 27} (1988) 587--611.

\item{\ 5.} {\gross A.\ J.\ Ellis}: {\sl Minimal Decompositions in
            Partially Ordered Normed Vector Spaces.} Proc.\
            Camb.\ Phil.\ Soc.\  
            {\bf 64} (1968) 989--1000.

\item{\ 6.} {\gross H.\ H.\ Schaefer}: {\sl Banach Lattices and 
            Positive Operators}. Springer, Berlin, 1974.

\item{\ 7.} {\gross E.\ Ruch, W.\ Stulpe}: {\sl The Proof of
            the Mixing Theorem for Statistical Systems in
            Classical Physics}. Preprint, 1997, to appear in
            Acta.\ Applic.\ Math.

\item{\ 8.} {\gross P.\ Busch, R.\ Quadt}: {\sl Operational
            Characterization of Irreversibility}. Preprint, 1997.

\item{\ 9.} {\gross P.\ Busch }: {\sl Stochastic Isometries in
            Quantum Mechanics.} Preprint, 1998.

\bye